\documentstyle[psfig,floats,aps] {revtex}
\begin{document}
\newcommand{\vel}{{\bf v}}

\twocolumn[\hsize\textwidth\columnwidth\hsize\csname@twocolumnfalse\endcsname
\author{Jonathan Miller$^1$ and Peter B.~Weichman$^2$}
\address {$^1$NEC Research Institute, 4 Independence Way, Princeton, NJ 08540 \\
$^2$Blackhawk Geometrics, 301 Commercial Rd., Suite B, Golden, CO 80401}
\title{Chern numbers and localization by non-Hermitean operators}
\date {{\today}}
\maketitle

\begin{abstract}
In Hermitean quantum mechanics, extended current-carrying states are
distinguished from localized ones by their non--zero Chern number.
We generalize the notion of Chern 
number to non-Hermitean localization problems such as tilted
flux lines in type-II superconductors with line defects and passive 
scalar transport in random vorticity fields.  We find that the usual 
point eigenvalue degeneracies that occur in the Hermitean case are now 
loops bounding a branch cut sheet with the same quantized total 
Berry flux, and hence Chern number, as the original point.  If a loop 
cuts the integration surface non-quantized contributions to the total 
flux result.
\end{abstract}
\pacs{PACS numbers ....}
\vskip2pc]
\narrowtext

In Hermitean quantum mechanics the differential geometric
notion of Chern number arises in the context of adiabatic transport
of eigenstates in Hilbert space~\cite{berry}. The change in the overall 
phase of a state after it is transported around a closed loop contains
a purely geometric contribution associated with curvature
of the Hilbert space. Quantized integer Chern numbers are associated 
with singularities in the curvature that originate in
eigenvalue degeneracies of the Hamiltonian generating the transport.

One way of transporting an eigenstate is by changing its boundary 
conditions.  A Chern number can then be computed as a certain
integral over the space of boundary conditions.  Since a 
localized state is insensitive to boundary conditions, a nonzero 
Chern number should be associated only with an extended state, 
and serves as a nontrivial diagnostic of localization.  This 
procedure has been used to study localization in the integer quantum
Hall effect~\cite{IQHE}. Physically, the Chern number reflects
the ability of the state to transport charge across the system, 
and is proportional to its Hall conductance~\cite{berry}.
 
Recently there has been much interest in localization properties
of certain \emph{non}-Hermitean operators, for example, operators
governing effective two-dimensional quantum descriptions
of pinning of tilted flux lines by extended defects in three-dimensional
type-II superconductors~\cite{HN}, and passive scalar transport by
two-dimensional random vorticity fields~\cite{MW}.  Both these examples
have close ties to the problem of electron localization in random
magnetic fields~\cite{rflux}.  In this 
paper we will consider the generalization of Chern numbers to non-Hermitean 
operators. We review the non-Hermitean Berry phase~\cite{GW88} and how it 
gives rise to Chern numbers through eigenvalue degeneracies. Application is
then made to a class of operators that interpolate, as a function
of a parameter $0 \leq \theta < \pi/2$, between a special case of
the Hermitean random flux model and the non-Hermitean flux line and passive 
scalar problems.

Consider a family ${\bf L}({\bf X})$ of (possibly non-Hermitean) 
operators, where ${\bf X} = (X_1,X_2,\ldots)$ is a set of fixed {\it real}
parameters spanning a manifold.  Let $|m;{\bf X} \rangle$ and 
$\langle m;{\bf X}|$ be the right and left eigenvectors of ${\bf L}({\bf X})$
satisfying
\begin{eqnarray}
{\bf L}({\bf X}) |m;{\bf X} \rangle &=& E_m({\bf X}) |m;{\bf X} \rangle
\nonumber \\
\langle m; {\bf X}| {\bf L}({\bf X}) &=& E_m({\bf X}) \langle m; {\bf X}|
\nonumber \\
\langle m; {\bf X}| n;{\bf X} \rangle &=& \delta_{mn}.
\label{1}
\end{eqnarray}
The eigenstates are determined up to an arbitrary complex amplitude,
and we make a fixed arbitrary choice of this quantity so that 
$|m;{\bf X} \rangle$ is a single-valued function of ${\bf X}$ 
over the manifold.  The projection operator,
${\bf P}_m({\bf X}) = |m;{\bf X} \rangle \langle m; {\bf X}|$, is 
independent of this amplitude.  The eigenvalues $E_m({\bf X})$, assumed 
nondegenerate for now, may be complex, and in general 
$\langle m; {\bf X}|^\dagger \neq |m;{\bf X} \rangle$.  For any given 
operator ${\bf L}(t)$, we define the dynamics of states $|\phi(t) \rangle$ 
and $\langle \phi(t)|$ via
\begin{equation}
i \partial_t |\phi(t) \rangle = {\bf L}(t) |\phi(t) \rangle,~~
-i \partial_t \langle \phi(t)| = \langle \phi(t)| {\bf L}(t).
\label{2}
\end{equation}
This dynamics preserves all inner products $\langle \phi_1(t)|\phi_2(t) \rangle$.

Let ${\bf X}(\alpha)$, $0 \leq \alpha \leq 1$, ${\bf X}(0) = {\bf X}(1)$, 
parameterize a closed path, $C$, in parameter space.  We define \emph{parallel
transported} eigenstates $|m;\alpha \rangle = b_m(\alpha) |m;{\bf X}(\alpha) 
\rangle$ along the path ${\bf X}(\alpha)$ by the rule 
\begin{equation}
\partial_\alpha b_m(\alpha) = - \langle m;{\bf X}(\alpha)| 
\partial_\alpha |m; {\bf X}(\alpha) \rangle b_m(\alpha),
\label{3}
\end{equation}
with $b_m(0) = 1$.  The solution is~\cite{GW88}
\begin{eqnarray}
b_m(\alpha) &=& \exp \left[i\int_{C:{\bf X}(0)}^{{\bf X}(\alpha)} 
{\bf A}_m({\bf X}) \cdot d{\bf X} \right] \nonumber \\
{\bf A}_m({\bf X}) &\equiv& \langle m;{\bf X}(\alpha)| 
i\partial_{\bf X} |m; {\bf X}(\alpha) \rangle.
\label{4}
\end{eqnarray}
Direct substitution shows that this parallel transport 
may be generated dynamically using the operator ${\bf L}(t) 
= i[\partial_t{\bf P}_m({\bf X}(t)),{\bf P}_m({\bf X}(t))]$ 
(with $\alpha = t$)~\cite{adiabatic}. For a closed loop one obtains
\begin{equation}
b_m(1) = e^{i\gamma_m(C)},~~\gamma_m(C) 
= \int_C d{\bf X} \cdot {\bf A}_m({\bf X}).
\label{5}
\end{equation}
In the Hermitean case ${\bf A}_m$, and hence $\gamma(C)$, is real.  In the 
non-Hermitean case they are both generally complex.  It is easy to see
that $\gamma(C)$ is independent of the arbitrary normalization of the states
$|m;{\bf X} \rangle$.  Any change of normalization, $e^{i\lambda({\bf X})}$, 
appears as a gauge transformation ${\bf A}^\prime = {\bf A} 
- \partial_{\bf X} \lambda$, under which the integral in (\ref{5}) is 
invariant. Provided no singularities are encountered we may
write $\gamma(C)$ as the integral of the gauge invariant
flux using Stokes theorem,
\begin{eqnarray}
{\bf F}_m({\bf X}) &=& \nabla \times {\bf A}_m({\bf X}) \nonumber \\
&=& i [\partial_{\bf X} \langle m_0;{\bf X}|] 
\times [\partial_{\bf X} |m_0;{\bf X} \rangle], 
\label{6}
\end{eqnarray}
over any surface $S$ bounded by $C$~\cite{foot2}:
\begin{equation}
\gamma_m(C) = \int_S d\Sigma \cdot {\bf F}_m({\bf X}).
\label{7}
\end{equation}
By construction one has $\nabla \cdot {\bf F}_m = 0$, as is required for
$\gamma(C)$ to be independent of the choice of $S$.

By inserting a complete set of states, the flux may be written 
in the form
\begin{equation}
{\bf F}_{m_0}({\bf X}) = i \sum_n{}^\prime {\langle m; {\bf X}|
\partial_{\bf X} {\bf L} |n;{\bf X} \rangle \times
\langle n; {\bf X}| \partial_{\bf X} {\bf L} |m;{\bf X} \rangle
\over (E_n-E_m)^2}
\label{8}
\end{equation}
where the prime indicates that the term $n=m$ is omitted.
Equation (\ref{8}) demonstrates explicitly that so long as 
${\bf L}({\bf X})$ is smooth, singularities in ${\bf F}_m$
can only occur when two eigenvalues cross:  $E_n - E_m \to 0$
for some $n \neq m$. In the Hermitean case eigenvalue crossings,
and therefore singularities in ${\bf F}_m({\bf X})$, generically occur 
at isolated points in the three dimensional space of ${\bf X}$.  
One may choose the surface $S$ in (\ref{7}) to avoid such 
points by passing either over or under them.  On the one hand,
the difference of the integrals over two such
surfaces $S_1$ and $S_2$ is
\begin{equation}
\gamma_m(S) \equiv \gamma_m(S_1) - \gamma_m(S_2) 
= \int_S d\Sigma \cdot {\bf F}_m,
\label{10}
\end{equation}
where $S = S_1 \cup S_2$ is a closed surface enclosing
the singularity. On the other hand, $e^{i\gamma_m(S_1)}
= e^{i\gamma_m(S_2)}$ since both surfaces are bounded by 
the same curve $C$.  We conclude that for any closed surface $S$, 
$\gamma_m(S) = 2\pi q$ for some integer {\it Chern number} $q$~\cite{foot1}.  
Eigenvalue degeneracies act as a quantized point sources of flux 
for ${\bf F}_m$. 

This argument does not rely directly on Hermiticity,
but only on the result that energy level degeneracies
occur generically at points in three dimensional space, so that
these singularities can be enclosed by singularity-free
closed surfaces.  We now address the issue of how to generalize
this result to the non-Hermitean case.
In order to study a degeneracy in more detail, we project
${\bf L}$ onto the (complex) two-dimensional subspace spanned by the 
eigenvectors $|m;{\bf X}\rangle$ and $|n;{\bf X}\rangle$.
If ${\bf P}_{mn} = |m\rangle \langle m| + |n \rangle
\langle n|$ is the corresponding orthogonal projection, we
consider the matrix
\begin{eqnarray}
{\bf L}_{mn}({\bf X}) &=& {\bf P}_{mn}({\bf X}) {\bf L}({\bf X})
{\bf P}_{mn}({\bf X}) \nonumber \\
&=& \alpha_0({\bf X}) {\bf I} + {\vec \alpha}({\bf X}) 
\cdot {\vec \sigma}
\label{11} 
\end{eqnarray}
where $\vec \sigma$ are the Pauli matrices, and $\alpha_0$
and $\vec \alpha = (\alpha_1,\alpha_2,\alpha_3)$ are four
(at this stage arbitrary) complex numbers.  In the Hermitean
case these numbers are all real.  The eigenvalues of ${\bf L}_{mn}$ 
are $E_\pm = \alpha_0 \pm \Delta E$, where $\Delta E = 
\sqrt{{\vec \alpha} \cdot {\vec \alpha}}$.  The left and right
eigenvectors are easily computed as well, and the the fluxes are
\begin{equation}
{\bf F}_\pm({\bf X}) 
= \mp {\alpha_1 {\bf a}_2 \times {\bf a}_3 + \alpha_2 {\bf a}_3 \times
{\bf a}_1 + \alpha_3 {\bf a}_1 \times {\bf a}_2 \over 2 \Delta E^3},
\label{14}
\end{equation}
where ${\bf a}_i({\bf X}) = \partial_{\bf X} \alpha_i({\bf X})$.

Equation (\ref{14}) can be used to explore the nature of the 
singularities in ${\bf F}_\pm$ in the neighborhood of degeneracies, 
$\Delta E \to 0$.  The condition for $\Delta E=0$ is simply $\alpha_1^2 
+ \alpha_2^2 + \alpha_3^2 = 0$.  In the Hermitean case where the 
$\alpha_i$ are all real, this condition requires that $\alpha_1 
= \alpha_2 = \alpha_3 = 0$, but in general this equality places 
only {\it two} conditions on the six parameters (the real and 
imaginary parts of each $\alpha_i$, $i=1,2,3$).  In the parameter 
space ${\bf X}$ degeneracies will then generically occur 
on a submanifold of codimension 2, i.e., on {\it lines} in the three 
dimensional space.  

Since the matrix ${\bf L}_{mn}$ is no longer Hermitean, there is no 
guarantee that there will be two independent eigenvectors associated 
with the two degenerate eigenvalues.  Generically ${\bf L}_{mn}$ 
will be upper--triangular (with $\alpha_1 + i \alpha_2 = 0$ at the same
time that $\alpha_3 = 0$, but $\alpha_1 - i \alpha_2 \neq 0$), and 
only one eigenvector will exist, with the orthogonal direction 
spanned by a {\it generalized eigenvector}~\cite{foot3}.  We shall 
call this situation a {\it generalized} degeneracy.  What we call 
a {\it true} degeneracy exists only when both $\alpha_1 \pm i\alpha_2 
= 0$, so that ${\bf L}_{mn} = E{\bf I}$ is proportional to the identity
matrix (this differs from the Hermitean case only in that $\alpha_0
= E$ may be complex), and places {\it six} conditions on the
occurence of a true degeneracy, which will generically {\it not} 
occur for a three dimensional parameter space ${\bf X}$, unless further
restrictions are placed on the class of matrices (such as 
self-adjointness). In what follows we do not impose any such further
restrictions.

If ${\bf X}_g$ be a point of generalized degeneracy, set
${\bf x} = {\bf X} - {\bf X}_g$.  From (\ref{14}) we obtain for
small ${\bf x}$:
\begin{equation}
{\bf F}_\pm = \mp {{\bf v} \over 2^{5/2} ({\bf w} \cdot {\bf x})^{3/2}},
\label{15}
\end{equation}
with complex vectors
\begin{eqnarray}
{\bf v} &=& \alpha_1 {\bf a}_2 \times {\bf a}_3 + \alpha_2 {\bf a}_3 
\times {\bf a}_1 + \alpha_3 {\bf a}_1 \times {\bf a}_2 \nonumber \\
{\bf w} &=& \alpha_1 {\bf a}_1 + \alpha_2 {\bf a}_2 + \alpha_3 {\bf a}_3,
\label{16}
\end{eqnarray}
with $\alpha_1^2 + \alpha_2^2 + \alpha_3^2 = 0$.
The easily proven result ${\bf v} \cdot {\bf w} = 0$ guarantees that
$\nabla \cdot {\bf F}_\pm = 0$.  

The condition ${\bf w} \cdot {\bf x} 
= 0$ is satisfied locally if ${\bf x}$ is orthogonal to {\it both}
${\bf w}_R \equiv {\rm Re} {\bf w}$ and ${\bf w}_I \equiv {\rm Im} 
{\bf w}$, showing that the line of generalized degeneracies is
locally tangent to ${\bf u} = {\bf w}_R \times {\bf w}_I$.  Note that
unless it is real, ${\bf v}$ will not be parallel to ${\bf u}$. 
This line can pierce the integration surface $S$ at singular points 
where ${\bf F}_\pm$ will have the asymptotic behavior (\ref{15}).  
Since the surface integral is two dimensional, the singularity will 
be {\it integrable}.

Suppose now that ${\bf x}$ circles a singular line in the plane
formed by ${\bf w}_R$ and ${\bf w}_I$.  For example, let 
${\bf w} \cdot {\bf x} = \eta e^{i\theta}$ with $\eta$ fixed and 
$-\pi < \theta \leq \pi$ determined by ${\bf w}_R \cdot {\bf x} 
= \eta \cos(\theta)$ and ${\bf w}_I \cdot {\bf x} = \eta \sin(\theta)$.
Then ${\bf F}_\pm \propto e^{-i3\theta/2}$, so that single valuedness
can be maintained only if a {\it surface of branch cuts} extends out of
the singular line.  It is straightforward to show that this branch sheet 
joins ${\bf F}_+$ and ${\bf F}_-$:  tracing the closed path causes 
$E_m$ and $E_n$ to exchange positions by circling each other in the 
complex plane.  ${\bf F}_\pm({\bf X})$ forms a
single complex function on a two-sheeted Riemann surface 
(see Fig.~\ref{fig2} below).

In the Hermitean case one may deform the integration
surface to small spheres enclosing the point singularities.  In
the generic case the deformed surface wraps around those portions of
the lines and their branch surfaces that are internal to the original
surface.

In order to understand the result of the integration (\ref{7}) 
we consider a model problem with the {\it exact}
form ${\vec \alpha} = (\alpha_1^0 + {\bf a}_1 \cdot {\bf x},
\alpha_2^0 + {\bf a}_2 \cdot {\bf x},\alpha_3^0 + {\bf a}_3 
\cdot {\bf x})$, so that a true degeneracy exists at ${\bf x}={\bf 0}$ 
when $\vec \alpha^0 = (\alpha_1^0,\alpha_2^0, \alpha_3^0) = {\bf 0}$.
Near ${\bf 0}$ the flux is
\begin{equation}
{\bf F}_\pm({\bf x}) = \mp {D_0 {\bf x} \over 2 ({\bf S}{\bf x} \cdot 
{\bf S}{\bf x})^{3/2}}
\label{17}
\end{equation}
where ${\bf S}$ is the $3 \times 3$ matrix whose rows are the 
${\bf a}_i$, and $D_0 = \det{\bf S} = {\bf a}_1 \cdot {\bf a}_2 
\times {\bf a}_3$.

We suppose that the ${\bf a}_i$ are all real, so 
that $\vec \alpha^0$ represents a (generally non-Hermitean) perturbation 
of an {\it Hermitean} problem.  On transformation to real coordinates
${\bf y} = {\bf S}{\bf x}$, the flux takes the classic Coulomb form
${\bf F}_\pm({\bf y}) = \mp \sigma(D_0) {\bf y}/2|{\bf y}|^3$,
where $\sigma(D_0)$ is the sign of $D_0$. It follows immediately
that the integrated flux through any surface $S$ enclosing the origin is
$\gamma(S) = \mp 2\pi \sigma(D_0)$.

On including the constant shift $\vec \alpha^0$, the flux becomes
\begin{equation}
{\bf F}_\pm({\bf y}) = \mp \sigma(D_0) {\vec \alpha^0
+ {\bf y}  \over 2 [(\vec \alpha_0 + {\bf y}) \cdot 
(\vec \alpha_0 + {\bf y})]^{3/2}}.
\label{19}
\end{equation}
For very large $|{\bf y}|$ this form reduces to (\ref{17}).  The
flux through a surface at infinity remains $\gamma(S)$.
Since $\nabla_{\bf y} \cdot {\bf F}_\pm({\bf y}) = 0$,
this result holds for any surface that
encloses all singularities in ${\bf F}_\pm$.  These singularities
occur at the zeros of $(\vec \alpha^0 + {\bf y})^2$, i.e., for
$|{\bf y} + {\rm Re} \vec \alpha^0|^2 = |{\rm Im} \vec \alpha^0|^2$
and ${\rm Im} \vec \alpha^0 \cdot ({\bf y} + {\rm Re} \vec \alpha^0) = 0$,
which constitute a circular ring of radius $a = |{\rm Im} \vec \alpha^0|$ 
formed by the intersection of the sphere of radius $a$ centered 
at $-{\rm Re}\vec \alpha^0$ with the plane normal to ${\rm Im} 
\vec \alpha^0$ passing through this same center.  In ${\bf x}$-space
the ring becomes an ellipse. A ring must be spanned by a branch
surface in order to maintain single--valuedness of ${\bf F}_\pm$.

We have therefore established that under non-Hermitean perturbations
the usual quantized point charges expand into closed loops, with 
net flux quantized by the original Hermitean Chern number. If a loop
grows to pierce the surface of integration, only a fraction of
the total flux will be enclosed, and integer quantization 
is lost.  This fraction depends on the geometry of the branch cut,
which is itself arbitrary, except for its end points.
Only the {\it total} flux, integrated over 
{\it all} sheets of the Riemann surface, remains quantized~\cite{foot5}.
The latter will correspond to the net Chern number of the colliding 
eigenvalues, which are now topologically ``entangled.'' 
The entanglement becomes more pronounced as the
non-Hermiticity increases, and may involve more
than two eigenvalues that collide with each other sequentially as
${\bf X}$ is varied, generating further Riemann sheets.

We illustrate this formal discussion by a specific example,
a family of operators in two dimensions:
\begin{equation}
{\bf L} = - e^{i\theta} D \nabla^2 + i {\bf A}({\bf r}) \cdot \nabla,
\label{21}
\end{equation}
in which $0 \leq \theta < 2\pi$ controls the degree of non-Hermiticity
(${\bf L}$ is Hermitean only for $\theta = 0,\pi$), $D$ is a constant
coefficient, and ${\bf A}$ is an incompressible vector field:  $\nabla 
\cdot {\bf A} = 0$.  For $\theta = 0$, this operator corresponds to the 
random flux problem in the Coulomb gauge~\cite{rflux}, with magnetic field 
$B = \nabla \times {\bf A}$, and a particular choice $V = -{\bf A}^2/4D$ 
for the scalar potential.  For $\theta = \pi/2$ this operator represents
a model of flux lines in a three-dimensional superconductor with 
extended defects, (where ${\bf A}$ corresponds to the horizontal 
components of the magnetic field~\cite{HN}), or a two-dimensional
passive-scalar transport model, (where ${\bf A}$ is the
velocity field, and $D$ is the diffusion constant).  We set $D=1$ and 
implement (\ref{21}) numerically by discretizing it~\cite{MW} on a small 
($4 \times 4$) lattice. The components of the vector field ${\bf A}$ are 
initially chosen to be independent random variables chosen uniformly on 
an interval $[-W,W]$, with $W=40$.  Incompressibility is imposed by 
subtracting from ${\bf A}$ the function $\nabla a$, where $a({\bf r})$ 
satisfies $\nabla^2 a = \nabla \cdot {\bf A}$ with periodic boundary 
conditions.

Study of random operators like (\ref{21}) is motivated by questions of
localization~\cite{IQHE,HN,MW,rflux}. 
Extended states are distinguished from localized states by their sensitivity to 
boundary conditions. Let $E_n(\phi_x,\phi_y)$, $\psi_n^{\phi_x,\phi_y}(x,y)$ 
be the eigenvalues and eigenstates of ${\bf L}$ on an $L_x \times L_y$ system with 
boundary conditions $\psi_n^{\phi_x,\phi_y}(x+L_x,y) 
= e^{i\phi_x}\psi_n^{\phi_x,\phi_y}(x,y)$, $\psi_n^{\phi_x,\phi_y}(x,y+L_y) 
= e^{i\phi_y}\psi_n^{\phi_x,\phi_y}(x,y)$.  The two--dimensional space 
$(\phi_x,\phi_y)$ has the geometry of a torus that is
embedded in a three--dimensional parameter space ${\bf X}.$
The Berry flux through the torus is the integrand of (\ref{10}):
$F_n(\phi_x,\phi_y) = \langle \partial_{\phi_x} \psi_n
| \partial_{\phi_y} \psi_n \rangle 
- \langle \partial_{\phi_y} \psi_n | \partial_{\phi_x} 
\psi_n \rangle$, and if the integral of the flux
over the surface (the Chern number) is non-zero, then
the eigenstate is extended.

\begin{figure}
\begin{picture}(-400,20)
\put(0,-340){\special{hscale=50.5 vscale=50.5 psfile=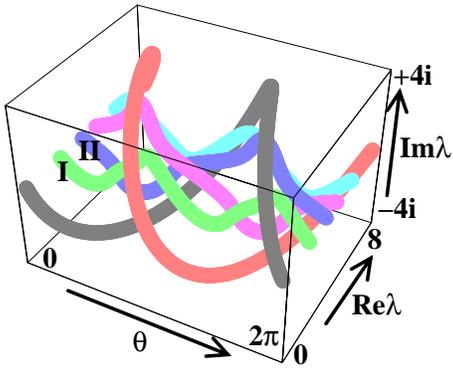}}
\end{picture}
\vskip1.75in
\caption{The middle six eigenvalues $\lambda$ as a function of
non-Hermiticity parameter $\theta$ for $\phi_x=\phi_y=0$.}
\label{fig1}
\end{figure}

Fig.~\ref{fig1} displays the variation of eigenvalues with
$\theta$ for $\phi_x=\phi_y=0$.  In Fig.~\ref{fig2} we plot the 
imaginary parts of $F_\pm(\phi_x,\phi_y)$ for the labeled
eigenvalues in Fig.~\ref{fig1} and for two values of $\theta$,
revealing the emergence of a branch cut and showing $F_\pm$ are 
different sheets of the same analytic function. The log--log plot 
exhibits the 3/2-law (\ref{15}) as an endpoint of a branch 
cut is approached.

The action corresponding to (\ref{21})
at $\theta = \pi/2$ is the Wick rotation 
of the action at $\theta = 0$. A natural 
question is whether this analytic continuation to 
imaginary time is accompanied by a smooth variation in
properties of eigenstates, or whether a phase 
transition, e.g., metal to insulator, occurs as a 
function of $\theta$ (see ref.~\cite{MW}).
For finite $\theta$, we have shown that
the geometry of eigenvalue degeneracies can change dramatically,
but the notion of Chern number is still relevant.
Whether the entanglement of eigenvalues can
lead to a phase transition in the thermodynamic limit
is unknown.

In summary, we extend the notion of Chern
number to non-Hermitean operators and find that it remains a useful
tool for distinguishing extended and localized states.  
Entanglement of eigenstates can lead to an intricate multi-sheeted
analytic structure for the Berry flux, and the Chern number becomes
defined only for the group of entangled states as a whole. 

\begin{figure}
\begin{picture}(-400,-20)

\put(-70,-385){\special{hscale=53.5 vscale=53.5 psfile=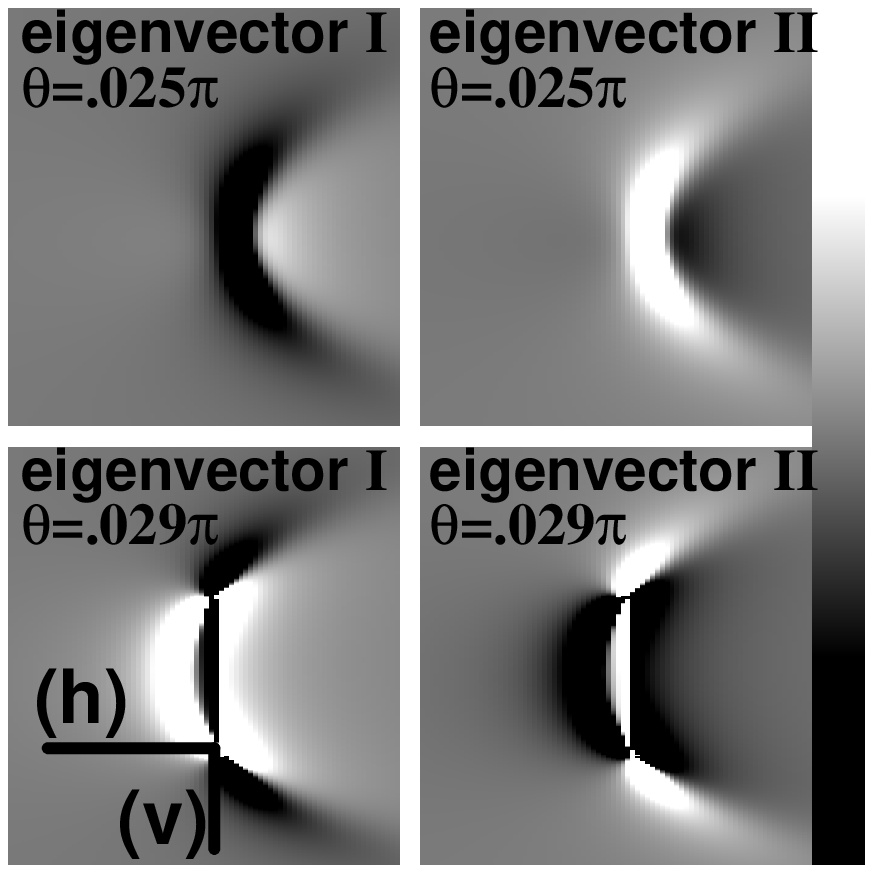}}
\put(135,-160){\special{hscale=20.5 vscale=22.5 psfile=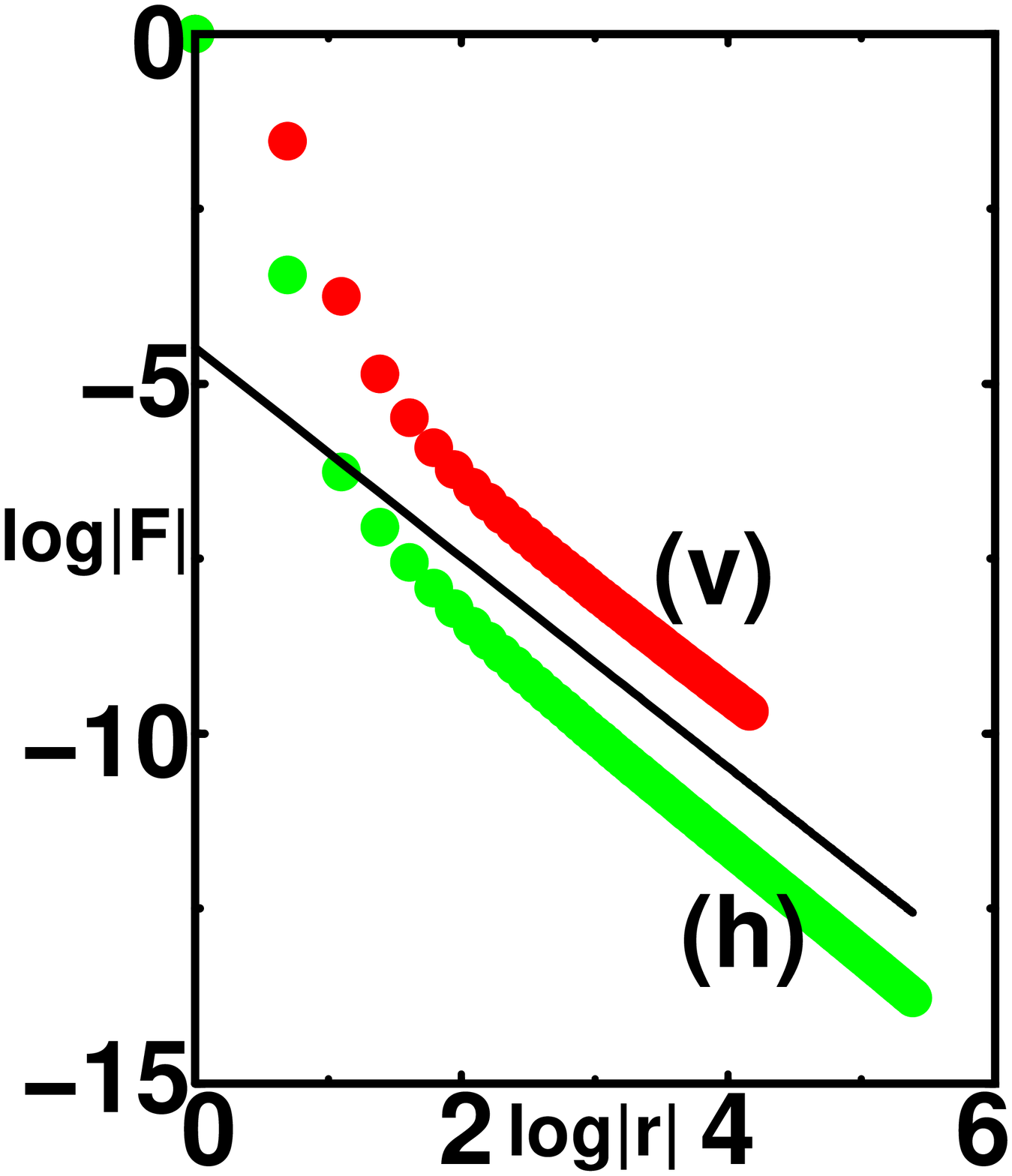}}

\end{picture}
\vskip2.05in
\caption{Plot of imaginary part of the Berry flux Im$F_\pm(\phi_x,\phi_y)$ 
for two values of $\theta$, in a small part of the Brillouin zone, 
$[0.35\pi,0.40\pi] \times [1.24\pi,1.25\pi]$.
The Chern numbers for eigenvalues I and II are $+1$ and $-1$ respectively 
until the first branch cut (not shown) occurs at about $\theta=.021\pi$.
As shown on the left, another branch cut (the obvious vertical lines 
in the two lower panels separating white from black) emerges between 
$\theta=.025\pi$ and $\theta=.029\pi$.  The linear grey scale spans the
range $[-1.3,1.3]$.  The log--log plot on the right displays the divergence 
of the magnitude of the flux against the distance $r$ from an endpoint 
of the branch cut along the paths (h) and (v); the solid line has slope 
-3/2; deviations from it at small $r$ arise from finite numerical 
resolution.}
\label{fig2}
\end{figure}


\begin{references}
\bibitem{foot5} To prove this, observe that the Berry flux through
a branch surface is identical for the two eigenvalues involved, but the 
surface normals are opposite. Branch surfaces make no
contribution and the total integral is conserved and equal to the
sum of the Chern numbers before the Chern loop pierced the 
integration surfaces.
\end{references}
\end{document}